\documentclass[11pt,letterpaper]{article}
\usepackage[top=0.85in,left=1.75in,footskip=0.75in,marginparwidth=2in]{geometry}

\usepackage[utf8]{inputenc}

\usepackage{nameref}

\usepackage[right]{lineno}

\usepackage{microtype}
\DisableLigatures[f]{encoding = *, family = * }

\raggedright
\usepackage{setspace} 
\usepackage{changepage}

\usepackage[aboveskip=1pt,labelfont=bf,labelsep=period,singlelinecheck=off]{caption}

\usepackage[sorting=none]{biblatex}

\usepackage{textcomp}

\usepackage{lastpage,fancyhdr,graphicx}
\usepackage{epstopdf}
\fancyheadoffset[L]{0.5in}
\fancyfootoffset[L]{0.5in}

\usepackage{color}

\definecolor{Gray}{gray}{.25}

\usepackage{graphicx}

\usepackage{sidecap}

\usepackage{wrapfig}
\usepackage[pscoord]{eso-pic}
\usepackage[fulladjust]{marginnote}
\reversemarginpar

\begin{document}
\vspace*{0.35in}

{\Large
\justify{\bf{2D-MRI of the Central Nervous System: The effect of a deep learning-based reconstruction pipeline on the overall image quality}
}}
\newline

\begin{flushleft}

D.E. Gkotsis$^{1\dagger}$,
A. Vlachopoulou\textsuperscript{1},
K. Dimos\textsuperscript{2},
I. Seimenis\textsuperscript{3},
E. Despotopoulos\textsuperscript{2},
E.Z. Kapsalaki\textsuperscript{1,4}

\bigskip
$^{1}$ Institute Euromedica-Encephalos, Department of Magnetic Resonance, Athens, Greece
\\
$^{2}$ Institute Euromedica-Encephalos, Department of Computed Tomography, Athens, Greece
\\
$^{3}$ National and Kapodistrian University of Athens, School of Medicine, Department of Medical Physics, Athens, Greece
\\
$^{4}$ University of Thessaly, School of Medicine, Department of Radiology, Larissa, Greece
\\
\bigskip
$ ^{\dagger} $ \bf{Correspondence to: d.gkotsis@med.uoa.gr}

\end{flushleft}

\section*{Abstract}

\justify{Purpose of this study was to evaluate the effect of a robust magnetic resonance reconstruction pipeline equipped with a deep convolutional neural network on the overall image quality, in terms of Gibbs artifact reduction, and SNR improvement. Sixteen (16) healthy volunteers enrolled in this study and were imaged at 3T. Representative images of each image series that were reconstructed through the pipeline that leverages a deep learning (DL) algorithm were retrospectively benchmarked against corresponding images reconstructed through a conventional pipeline. DL-reconstructed images showed significant SNR improvements compared to the corresponding conventionally reconstructed images. In addition to that, Gibbs artifacts were effectively eliminated, when the raw data were reconstructed through the DL pipeline. Gibbs artifact reduction was qualitatively assessed by two
experienced medical physicists and two experienced radiologists. DL-based reconstruction can lead to an SNR surplus which can be further invested into either higher spatial resolution and thinner slices, or into shorter scan times.}

\section*{Introduction}

\justify{The advantages of magnetic resonance imaging (MRI) compared to other imaging modalities have been known for decades (i.e., use of non-ionizing radiation, higher soft tissue contrast etc.). One might argue that the most significant downside to high resolution MRI is the long acquisition time necessary for acquiring high resolution clinical images with sufficient signal-to-noise ratio (SNR).\\
Furthermore, in clinical practice, where usually more than one pulse sequences (minimum five different ones) are routinely obtained, per anatomical region, the excess time required to acquire high resolution/high SNR images, often, renders the acquisition unattainable. The practitioner, therefore, often, decides to sacrifice a little bit of spatial resolution (either in-plane or through-plane) to reduce the total acquisition time, attempting to strike a fair compromise between resolution, SNR, and total scan times.\\
In addition to that, spurious artifacts might sometimes arise, as a direct result of this exact compromise. For example, Gibbs artifacts manifest themselves in both k-space and image space, when the spatial resolution is low. Gibbs artifacts, also known as ringing artifacts, can be eliminated if the practitioner increases the spatial resolution, at the expense of a lower SNR and larger scan times.\\
In recent years, the field of artificial intelligence (AI) has seamlessly integrated with the field of medical imaging and has already yielded astonishing results. For example, Gupta et al \cite{1}, and Wang et al \cite{2} managed to employ a deep learning model and successfully generated synthetic sets of CT images from sets of MRI scans. Chen et al \cite{3} trained a three-dimensional convolutional neural network (CNN) on images of both low and high dose of gadolinium (Gd) contrast enhancement agent and managed to generate synthetic full Gd dose images from images acquired at only the 1/10th of full dose. Detailed overviews of the ways AI has aided the field of medical imaging can be found here \cite{4,5,6,7,8}.\\
In this study, we utilized and evaluated a new DL-based approach to two-dimensional (2D) image reconstruction in MRI, which is commercially available as AirReconDL{\texttrademark} (GE Healthcare, Milwaukee, USA), and is offered as an additional purchase option from GE Healthcare, for MRI scanners with magnetic field strengths ranging from 1.5T to 7.0T. The AirReconDL{\texttrademark} reconstruction pipeline comprises a deep convolutional neural network (CNN), which was trained, validated, and tested using high resolution/high SNR and low resolution/low SNR pairs of k-space datasets of various anatomical regions. The main aim of the current study was to evaluate the performance of this DL-based reconstruction pipeline, in terms of image quality characteristics (e.g., SNR and presence of truncation artifacts).}

\section*{Materials and Methods}

\justify{All sixteen (16) volunteers enrolled in this study have signed an informed consent form regarding the post-processing of the acquired imaging data and the potential publication of a paper including de-identified representative images of their data.\\
All images presented in this study were acquired using a 3.0T GE Signa Premier scanner, software version 29.0. Multi-channel phased-array coils were used for all studies. For brain scans, a 48-channel coil was used, whilst a 21-channel coil was used for cervical spine. For thoracic and lumbar spine scans, the anterior array 60-channel flexible coil was used.\\
The pulse sequences employed were fast spin echo (FSE), fast gradient recalled echo (FGRE), and inversion recovery (e.g., STIR, FLAIR) pulse sequences, with T$_{1}$-, T$_{2}$-, and T$_{2}$*-weighted contrast. Pulse sequences were used to image the following areas of the central nervous system: brain (21), cervical spine (11), thoracic spine (2), and lumbar spine (3), whilst they were prescribed in all orthogonal slice orientations (axial, sagittal and coronal). Some pulse sequences were run with chemical shift fat suppression. Details of the main acquisition and reconstruction parameters can be found in Table 1.\\
When reconstructing images through the DL-based reconstruction pipeline, the user is given the option to export the images that would be produced in the absence of the deep CNN embedded in the pipeline. These sets of images undergo the conventional reconstruction process and have identical acquisition parameters with the images that went through the DL-based reconstruction pipeline. Therefore, identical regions of interest (ROIs) were drawn on different tissues and in the background of corresponding images to estimate and compare signal to noise ratios (SNRs) achieved with the two reconstruction approaches. The SNR was estimated by dividing the average signal intensity of a tissue ROI with the average signal intensity of the background.}

\section*{Results}

\justify{As can be seen from Table 2, the largest SNR improvements were realized in the fast gradient echo pulse sequences, yielding a whopping 7-fold increase in SNR, compared to the same images reconstructed without the employment of the AirReconDL{\texttrademark}. \\
The average increase in SNR, regardless of coil used, pulse sequence type, acquisition parameters and type of contrast, across all anatomical areas covered and pulse sequences used was greater than 350\%.\\
Figure 1 displays representative brain images of a thin-slice coronal T$_{2}$ weighted fast spin echo pulse sequence, which was acquired with a small field of view of 14 cm. The image quality improved with DL-based reconstruction, and upon signal measurements, a 354\% increase in SNR was noted.\\
Representative spine images of coronal T$_{2}$ weighted fast spin echo pulse sequence with an additional inversion pulse to null the signal emanating from fat are shown in Figure 2. An SNR improvement of 388\% was realized.\\
Figure 3 depicts a T$_{2}$ weighted STIR image acquired with a matrix of 320$ \times $ 224 (top left) reconstructed with AirReconDL{\texttrademark} and the same image reconstructed conventionally (top right). The strength of the algorithm was set to “medium”. Upon acquiring and inspecting the image, we could not be sure if the line with the brighter signal was an artifact that resembles syringomyelia, or whether it was a finding. Therefore, we increased the resolution to 384x288, and set the strength of the algorithm to “high”, and increased the echo train length by 2, leading to a slight increase in acquisition time of 14 seconds, but with a final image of far superior image quality (bot left). This way, we were able to conclude that it was a normal finding, and neither an artifact nor a pathology. Same image reconstructed conventionally can be seen on the bot right. \\   
In Figures 4, 5, and 6, T$_{1}$-weighted FLAIR, T$_{2}$*-weighted FGRE, and T$_{1}$-weighted FSE pairs of images are shown, respectively. Corresponding SNR gains were 379\%, 680\%, and 229\%.\\
Figure 7 shows two representative images from a T$_{2}$ weighted fast spin echo pulse sequence. The arrows point to areas with visible Gibbs artifacts, when the final image was reconstructed through the conventional reconstruction pipeline (right), and the arrows on the left show their DL-reconstructed counterparts, where truncation artifacts are eliminated.\\
Since most of the pulse sequences acquired for this study were acquired at high spatial resolution, Gibbs artifacts were not easily observed (even in the conventionally reconstructed images). However, in all images that exhibited Gibbs artifacts, the algorithm essentially eliminated them entirely, thus elevating the overall image quality.
}

\section*{Discussion}

\justify{In general, deep learning is fundamentally different from conventional algorithms. In the latter, one must specifically define the various rules and relationships between variables, feed the input into the algorithm, and eventually, after the algorithm goes through all the rules and steps laid out by the programmer/developer, it yields an output. The output is then evaluated by the programmer/developer, and the troubleshooting process begins, by modifying the rules. On the other hand, in deep learning, the algorithm is fed both the inputs and the outputs and is then programmed to be able to discover the rules that connects the inputs and outputs by itself. \\
The training set of the convolutional neural network embedded in this reconstruction pipeline was further augmented through minor object displacements, either rotational, or linear. Furthermore, the image noise levels were enhanced through the addition of normally distributed noise. All the above, managed to yield an approximate number of two million unique pairs of images (four million images in total), allowing the algorithm to delineate the distribution and pattern of the noise inherent in the low SNR/low resolution images.\\
In machine learning in general, and in deep learning specifically, overfitting is one of the most serious issues that the developer must either overcome or avoid altogether. It has been shown that some of the most successful methods are the batch normalization \cite{9} and the use of dropout layers \cite{10}. In addition to these, usually there are multiple epochs involved in the training. When the algorithm goes through the entire dataset one time, this means that the model has been trained on one epoch. Usually, multiple epochs are being used, however, AirReconDL{\texttrademark} was trained in a single epoch. On top of the convolutional neural network residing in the reconstruction pipeline, there is also a generative adversarial network \cite{11}, which acts as a way of providing some sort of sanity-check feedback to the algorithm, preventing it from discovering false noise features, which might compromise its generalizability from the training set to the validation set and eventually to the testing set. A detailed technical description of the reconstruction pipeline and the deep CNN employed can be found here \cite{12}.\\
The DL-based reconstruction pipeline AirReconDL{\texttrademark} offers a tremendous increase in signal to noise ratio. This SNR surplus can be further invested into either improved image quality, or reduced scan times.\\
One limitation of this study, however, is the chosen method for SNR measurements. It has been demonstrated elsewhere \cite{13,14} that while ROI-based signal and noise measurements are the quickest, they are not the most robust. In addition to this, it has been shown that obtaining quantitative information regarding the noise power spectrum (NPS), yields the most accurate SNR measurements and estimations.\\
As mentioned earlier, the algorithm was trained on pairs of k-space images of low resolution/low SNR, and high resolution/high SNR. As a result, the algorithm managed to delineate the precise pattern of the Gibbs artifacts that commonly appear in images acquired at low resolution and/or with partial filling of the k-space, thus allowing the side-benefit of Gibbs artifact elimination that we demonstrated in Figure 7.
The results we obtained from this image quality inspection study, agree with the results of other investigators who have acquired and utilized this AI-based solution \cite{15,16,17}.\\
The seamless 2-fold (minimum) increase in SNR while at the same time achieving slightly higher spatial resolution, and slightly lower scan times, gives the user plenty of flexibility to further optimize the acquisition protocols. Furthermore, the fact that truncation/Gibbs artifacts are effectively removed, without having to resort to obscure resolution-degrading image filters, allows us to consider this reconstruction pipeline as a very powerful and indispensable tool. Furthermore, it has been shown that filters acting on the spatial domain, implement a pixel averaging, reducing thus the amount of noise in the image, causing however the effective spatial resolution to be also reduced.\\
A wide range of anatomic scans, pulse sequences and image contrast weightings are compatible and potentially improvable by utilizing AirReconDL{\texttrademark}. We believe the only downside to utilizing this reconstruction pipeline is incompatibility with motion-insensitive k-space filling techniques (i.e PROPELLER \cite{18} or BLADE \cite{19}). In cases where the patient is uncooperative and cannot stay still during the examination, we end up having to use a motion-insensitive pulse sequence, thus forfeiting the benefit of obtaining images with higher SNR/resolution.\\
Deep learning-based image reconstruction in MRI has proven to be an indispensable tool, which drastically improves image quality by effectively removing truncation artifacts, removing the background noise, without degrading the spatial resolution, since the algorithm is embedded in the reconstruction pipeline, and acts on the raw data of the k-space, in contrast to other types of denoising and/or sharpening filters that are commonly applied on the Fourier space data.}

\newpage

\section*{Acknowledgments}
\justify{The authors would like to acknowledge Dr. P. Athanasopoulos (GE Healthcare's Lead Clinical Applications Specialist for the MR modality in Southeastern Europe) for his valuable aid and thought-provoking discussions.}

\clearpage

\nolinenumbers

\medskip

\newpage

\printbibliography

\clearpage

\section*{Tables}

\centerline{\includegraphics{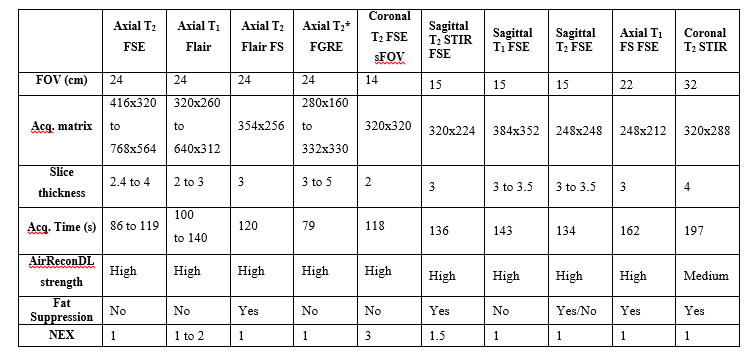}}
\textbf{Table 1.} Acquisition and reconstruction parameters of pulse sequences used.

\newpage

\centerline{\includegraphics{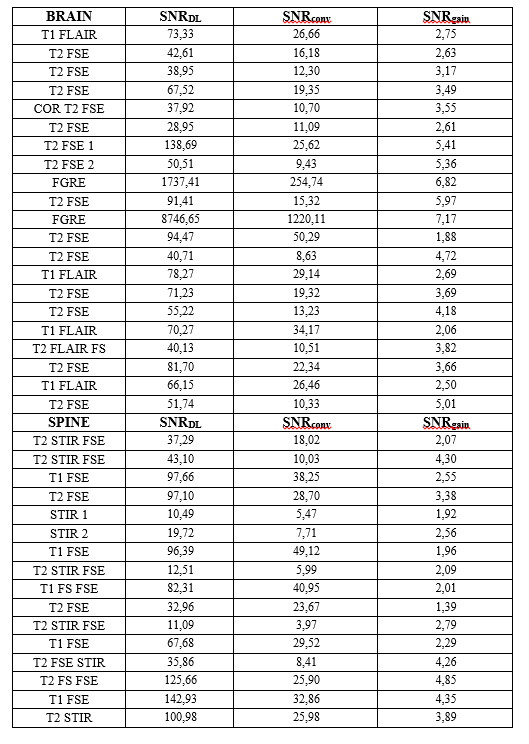}}

\justify{\textbf{Table 2.} Signal-to-noise ratios (SNRs) for images reconstructed with the DL-based algorithm (SNR$_{DL}$), and the conventional algorithm (SNR$_{conv}$), along with the respective SNR gains.}

\section*{Figures}

\centerline{\includegraphics[width=\textwidth,height=\textheight,keepaspectratio]{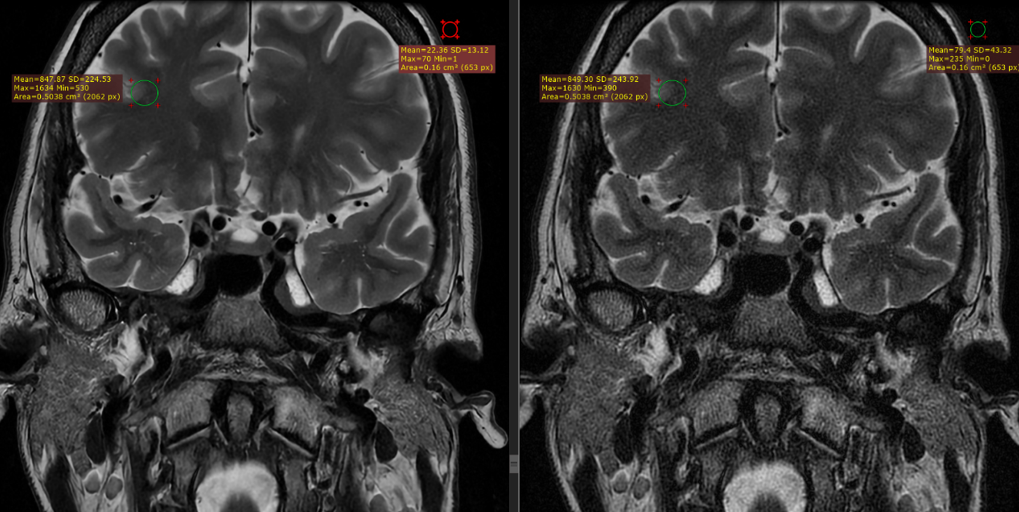}}

\justify{Figure 1. 2 mm coronal T$_{2}$ weighted fast spin echo pulse sequence, with a small field of view of 14 cm. SNR gains of 354\% were noted when the image was reconstructed through the DL-based reconstruction pipeline (left) compared to the conventional pipeline (right).}

\centerline{\includegraphics[width=\textwidth,height=\textheight,keepaspectratio]{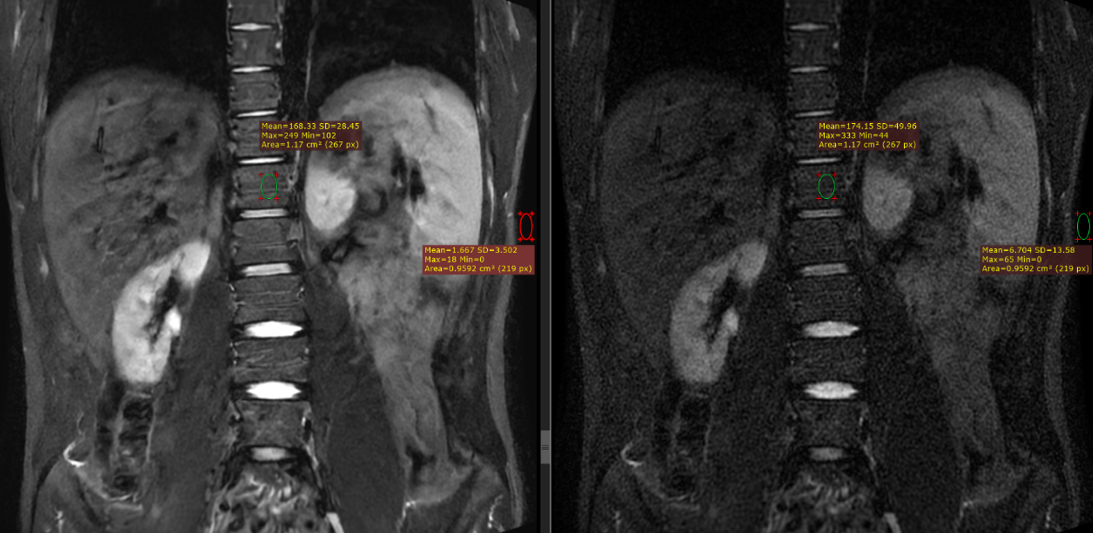}}
\justify{Figure 2. Coronal T$_{2}$ STIR fast spin echo image, acquired with a field of view of 32 cm. SNR gains of 388\% were noted, when the final image was reconstructed through the deep learning-based reconstruction pipeline (left), compared to the image reconstructed conventionally (right).}

\centerline{\includegraphics[width=\textwidth,height=\textheight,keepaspectratio]{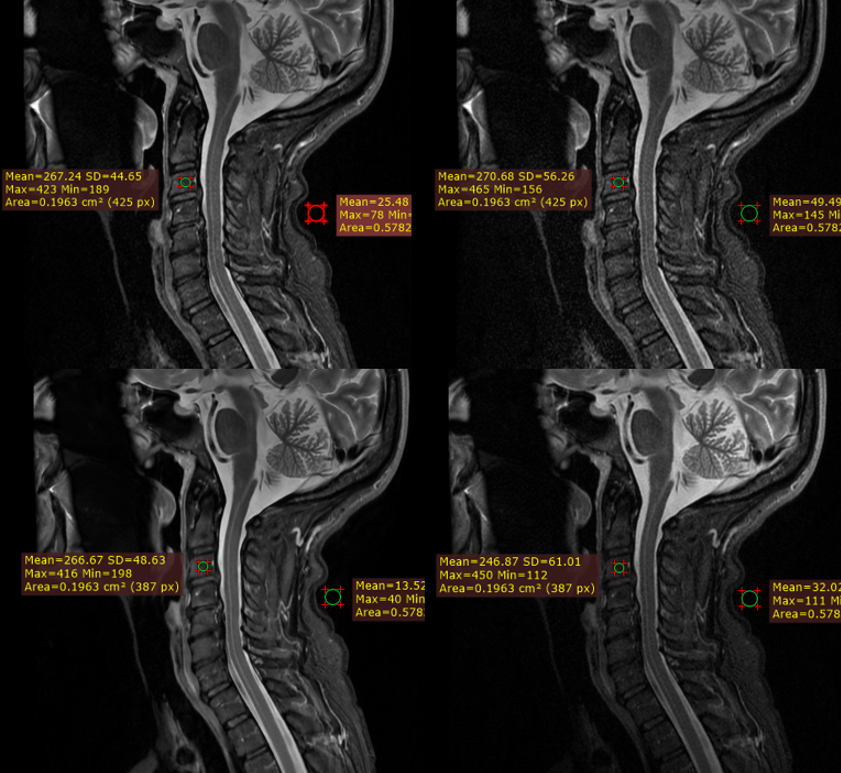}}

\justify{Figure 3. Sagittal fast spin echo STIR images. Top row has AirReconDL{\texttrademark} strength set to medium with an acquisition matrix of 320x224. Bottom row has AirReconDL{\texttrademark} strength set to high, and acquisition matrix of 364x288, increased echo-train length by 2. Images on the left are reconstructed through the DL-based pipeline, and images on the right are reconstructed conventionally. SNR gains were 192\% (top row), and 256\% (bot row).}

\centerline{\includegraphics[width=\textwidth,height=\textheight,keepaspectratio]{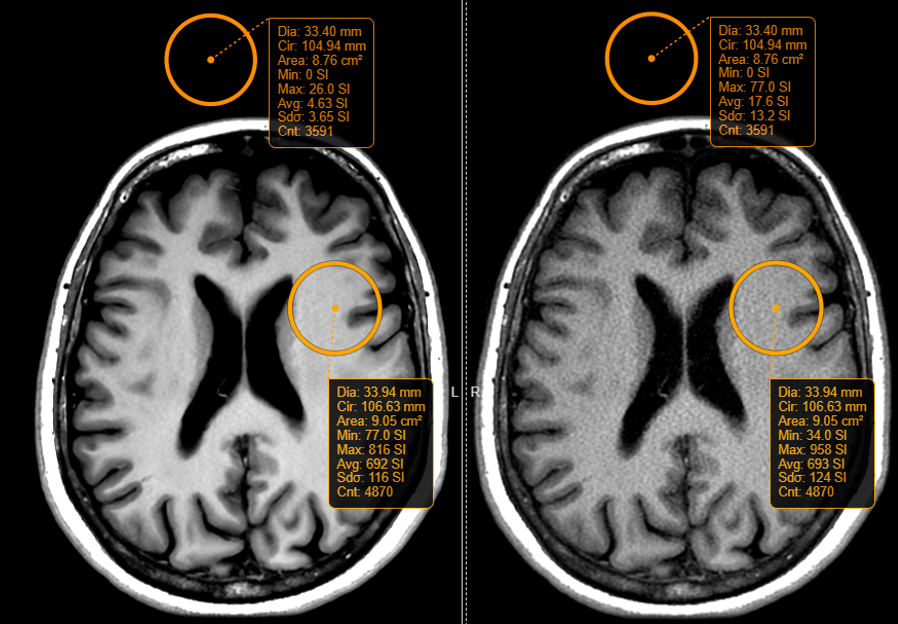}}

\justify{Figure 4. Axial images of 2mm slice thickness acquired with a T$_{1}$ fluid attenuation inversion recovery (FLAIR) pulse sequence. SNR gains of 379\% were noted when the image was reconstructed through the deep learning-based reconstruction pipeline (left) compared to the conventional pipeline (right).}

\centerline{\includegraphics[width=\textwidth,height=\textheight,keepaspectratio]{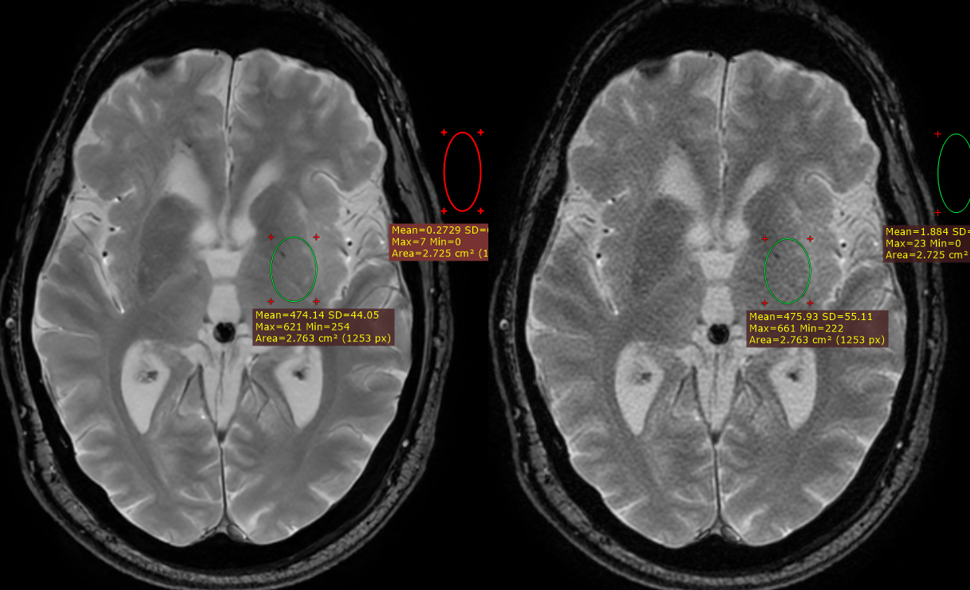}}

\justify{Figure 5. Axial images of 3 mm slice thickness acquired with a T$_{2}$* weighted fast gradient echo pulse sequence. SNR gains of approximately 680\% were noted when the image was reconstructed through the deep learning-based reconstruction pipeline (left) compared to the conventional pipeline (right).}

\centerline{\includegraphics[width=\textwidth,height=\textheight,keepaspectratio]{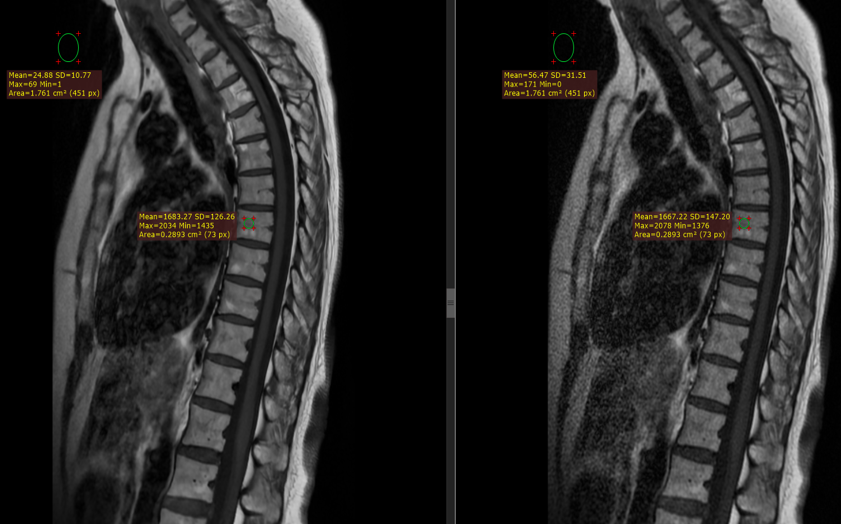}}

\justify{Figure 6. 3 mm thick sagittal T$_{1}$ weighted images acquired with a fast spin echo pulse sequence. SNR gains of approximately 229\% were noted when the image was reconstructed through the deep learning-based reconstruction pipeline (left) compared to the conventional pipeline (right).}

\centerline{\includegraphics[width=\textwidth,height=\textheight,keepaspectratio]{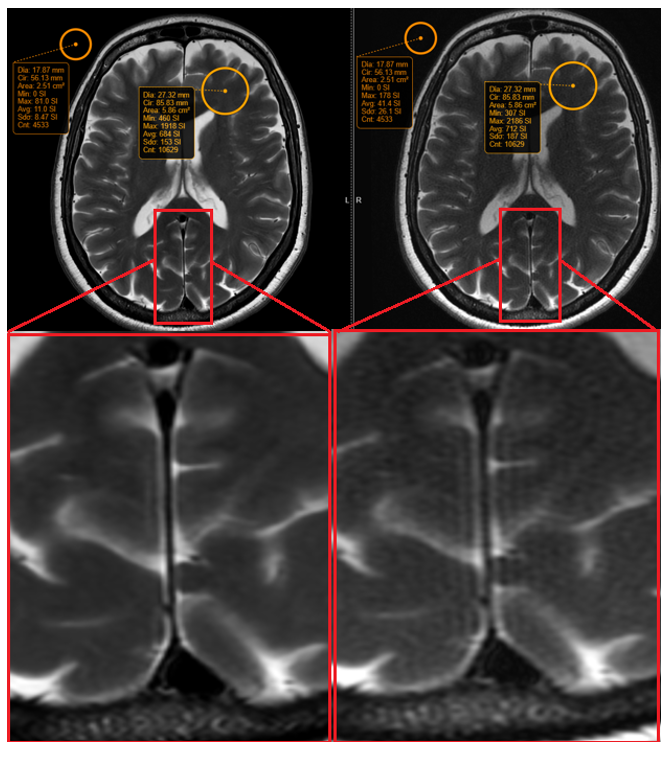}}

\justify{Figure 7. 3 mm thick sagittal T$_{2}$ weighted images acquired with a fast spin echo pulse sequence. SNR gains of approximately 362\% were noted when the image was reconstructed through the deep learning-based reconstruction pipeline (left) compared to the conventional pipeline (right). Note the Gibbs artifact elimination (red box).}

\end{document}